\documentclass[aps,superscriptaddress,prb,twocolumn,floatfix,showpacs,amsmath,amssymb]{revtex4}

\usepackage{color}
\usepackage{amssymb}
\usepackage{amsmath}

\usepackage{graphicx}
\usepackage[hypertex]{hyperref}

\begin{document}
\title{Constrain on superconductivity of potassium intercalated phenanthrene}

\author{Qiao-Wei Huang}
\affiliation{Department of Physics, South China University of Technology, Guangzhou 510640, China}
\affiliation{Center for High Pressure Science and Technology Advanced Research, Shanghai 201203, China}

\author{Guo-Hua Zhong}
\affiliation{Center for Photovoltaics and Solar Energy, Shenzhen Institutes of Advanced Technology, Chinese Academy of Sciences, Shenzhen 518055, China}

\author{Jiang Zhang}
\affiliation{Department of Physics, South China University of Technology, Guangzhou 510640, China}

\author{Xiao-Miao Zhao}
\affiliation{Department of Physics, South China University of Technology, Guangzhou 510640, China}
\affiliation{Center for High Pressure Science and Technology Advanced Research, Shanghai 201203, China}

\author{Chao Zhang}
\affiliation{Department of Physics, Yantai University, Yantai 264005, China}
\affiliation{Beijing Computational Science Research Center, Beijing 100084, China}

\author{Hai-Qing Lin}
\affiliation{Beijing Computational Science Research Center, Beijing 100084, China}
\affiliation{Center for Photovoltaics and Solar Energy, Shenzhen Institutes of Advanced Technology, Chinese Academy of Sciences, Shenzhen 518055, China}

\author{Xiao-Jia Chen}
\email{xjchen@ciw.edu}
\affiliation{Geophysical Laboratory, Carnegie Institution of Washington, Washington, DC 20015, USA}
\affiliation{Center for High Pressure Science and Technology Advanced Research, Shanghai 201203, China}

\date{\today}

\begin{abstract}

Raman-scattering measurements are performed in K${_x}$phenanthrene with the nominal value ${x}$ ranging from 0 to 6.0 at room temperature. Based on the Raman spectra, two phases (${x}$ = 3.0 and 4.0) are observed among the K${_x}$phenanthrene, and the immediate phase (${x}$ = 3.5) is produced. Only the ${x}$ = 3.0 phase is found to exhibit the superconducting transition at 5 K. The C-C stretching mods are observed to broaden and become disorder in K${_x}$phenanthrene (nominal value ${x}$ =2.0, 2.5, 6.0), indicating some molecular disorder in the metal intercalation process. This disorder is expected to influence the nonmetallic nature of these materials. The absence of metallic characteristics in those nonsuperconducting phases is provided from the calculated electronic structures based on the local density approximation.

\end{abstract}

\pacs{74.70.Kn, 74.62.Bf, 78.30.Jw, 74.20.Pq}

\maketitle

\section{INTRODUCTION}
Great interests are attracted on the studies of polycyclic aromatic hydrocarbons (PAHs) due to the recent discovery of superconductivity in these system. The PAHs compound, such as phenanthrene\cite{XFWang} (C$_{14}$H$_{10}$), picene\cite{picene} (C$_{22}$H$_{14}$), coronene\cite{Kubozono} (C$_{24}$H$_{12}$), and 1,2:8,9-dibenzopentacene\cite{GFChen} (C$_{30}$H$_{18}$), are found to exhibit superconductivity at temperatures from 5 to 33 K by doping alkali or alkali-earth metal. In contrast to the previous organic superconductors, such as tetrathiafulvalene (TTF),\cite{TTF} bis-ethylenedithrio-TTF(BEDT-TTF, abbreviated as ET),\cite{BEDTTF}  and tetramethyltetraselenafulvalene (TMTSF)\cite{TMTSF}  derivatives, hydrocarbon superconductors are novel superconducting materials with intriguing properties. Based on previous discoveries, $T_{\rm c}$ of hydrocarbon superconductor is enhanced with the increasing number of benzene rings,\cite{XFWang,picene,Kubozono,GFChen} reaching the highest $T_{\rm c}$ = 33.1 K in organic superconductos comparable to the carbon based superconductor metal-doped fullerides\cite{C60} ($T_{\rm c}$ = 38 K) under pressure. The feature of ``armchair" edge type, which is regarded as a key factor of superconductivity in such a system, is shared in the hydrocarbon superconductors in contrast to the ``zigzag" edge type.\cite{Kato,XFWang,picene,GFChen} Besides, hydrocarbon superconductors are also a low-dimensional system with strong electron-phonon and electron-electron interactions, and both of which may be involved in superconductivity as suggested by some theoretical works.\cite{Kato2,Subedi,Casula,Giovannetti,guo}

The mechanism of superconductivity in PAHs is still beyond understanding due to the complexity in such a system. Many theoretical works have been given some positive discussion in order to thoroughly understand the mechanism of superconductivity.\cite{T1,T2,T3} Rigid band approximation calculations on K$_{3.0}$picene show that the electrons transfer to the organic molecule from intercalated potassium atom and occupy its lowest unoccupied molecular orbital, leading to a metallic state. However, this approximation is not considered the influence of  electron-intramolecular-vibration interaction in K-doped picene, and the important electron-phonon interaction is experimentally evidenced by the photoemission spectroscopy.\cite{PS} Therefore, the charge transfer from the intercalated potassium atoms is expected to play an important role on the superconductivity of PAHs. Meanwhile, the importance of electron-intramolecular-vibration interaction can not be simply ignored in the investigation of the superconducting properties in this family.

Phenanthrene is the simplest molecule with an arm-chair conjuration that consists of three benzene rings. There is a larger superconducting volume fraction in cation doped phenanthrene compared to the other PAH based superconductors. \cite{XFWang,picene,Kubozono,GFChen}  The relationship between cation doping and superconductivity has not been established yet. It remains uncertain whether adding more cation element would be favor of superconductivity in this material.  Raman scattering is a useful technique to study the vibrational properties of the aromatic hydrocarbon molecules\cite{huang,zhao,fane,capi} and their doped derivative compounds.\cite{XFWang,picene,Kubozono,GFChen}  This technique has been proven very powerful in the determination of ${x}$ value in A${_x}$C$_{60}$ and A${_x}$picene.\cite{c1,c2,picene2} This method is a direct probe of the amount of the electrons those transferred to the molecule based on the shifts of the Raman modes. Previous investigations show that for A${_x}$C$_{60}$ and A$_{x}$picene, each electron contributes 6-7 cm$^{-1}$ in the Raman modes.

In this work we report  a detailed Raman scattering study of K${_x}$phenanthrene (0 $\leq$ ${x}$ (nominal) $\leq$ 6) at room temperature.  We determine the amount of electrons transferred to the phenanthrene molecule. Our results suggest the phase separation and intermediate phase in the nominal ${x}$ range from 0 to 6.0. Combining with density functional calculations, a constrain on superconductivity of potassium intercalated phenanthrene is obtained.

\section{EXPERIMENTAL DETAILS}

Phenanthrene of 98\% purity, colorless crystal, was purchased from Alfa Aesar, which was used without further purification. For each composition, phenanthrene powder was mixed with potassium at a nominal $x$ value in a glass tube, and the glass tube was sealed at 10$^{-4}$ par. Then, the sample was heated to the temperature at about 470 K for 40 hours. After the annealing process, the sample turns into uniform dark black color. The magnetic susceptibility and Raman spectrum measurements were conducted for the synthesized sample without any exposure to air. The magnetization $M$ was measured with the SQUID magnetometer (Quantum Design) at applied field $H$ of 10 Oe in a temperature range from 2.5 K to 8 K. Raman spectra were measured in backscattering geometry with visible laser excitation (633 nm) with power less than 5 mW at room temperature. All the samples were sealed in capillaries without the exposure to air.

\section{RESULTS AND DISCUSSION}

\begin{figure}[tbp]
\includegraphics[width=\columnwidth]{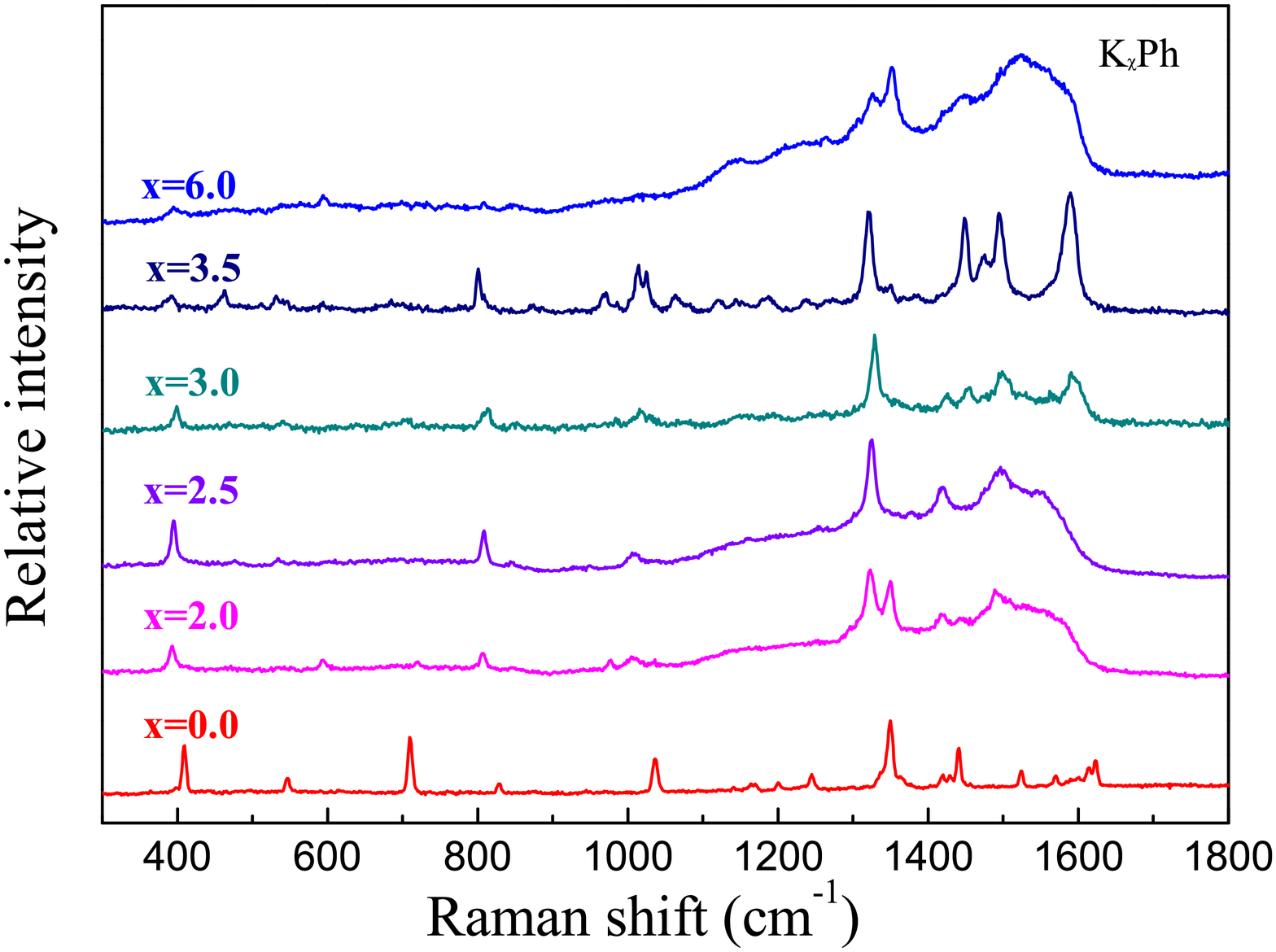}
\caption{(color online) Room temperature Raman spectra of K${_x}$phenanthrene (0 $\leq$ ${x}$ (nominal) $\leq$ 6).}
\end{figure}

Raman spectra of K${_x}$phenanthrene are shown in Fig. 1 with different ${x}$ value from 0 to 6.0. The spectrum of pristine phenanthrene agrees well with previously published data. There are four intense Raman peaks (originally at 409, 829, 1036, 1350 cm$^{-1}$) that can be made a comparison between pristine and metallic intercalated phenanthrene. All Raman modes of the K${_x}$phenanthrene obviously shift to the lower frequencies as the intercalation of potassium molecules. This indicates an Raman mode softening effect. The doping-induced downshift is pretty consistent with the previous studies of the similar materials, such as A${_x}$C$_{60}$ (A = K, Rb) and K${_x}$picene. These softening effects can be understood by the charge-transferred effect and the corresponding elongation of the intralayer C-C bond length due to the intercalation of alkali metal into these solids.

In the case of K${_{3.0}}$phenanthrene which exhibits superconductivity (shown in Fig. 2), there are 22 cm$^{-1}$ downshift for the 1350 cm$^{-1}$ mode. In the study of K$_{3.0}$phenanthrene,\cite{XFWang} superconductivity only occurs when the composition of alkali metal equals three. This phenomenon is also consistent with the studies of K${_x}$picene in which superconducting transition was observed when the intercalation of alkali metal closes to three. Consequently, it is expected to have three electrons transferred to the phenanthrene molecule in superconducting phase and thus each electron contributes to 6 or 7 cm$^{-1}$ redshift of the Raman modes at 1350 cm$^{-1}$.

\begin{figure}[tbp]
\includegraphics[width=\columnwidth]{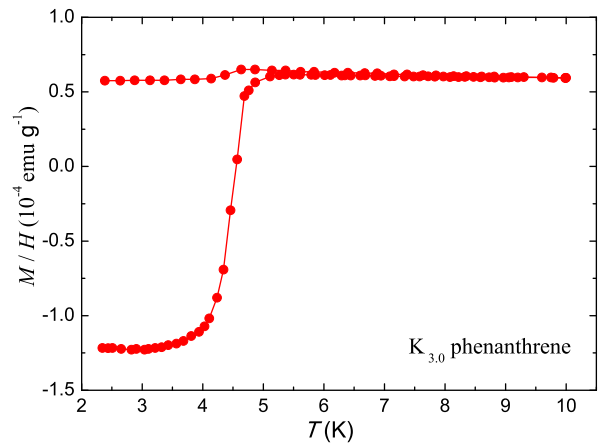}
\caption{ (color online) Temperature dependence of M/H for  K$_{3.0}$phenanthrene showing superconducting transition at 5 K in the ZFC and FC measurements at applied magnetic field of 10 Oe.}
\end{figure}

Actually, the redshift of the 409 and 1350 cm$^{-1}$ peak in K${_x}$phenanthrene is pretty consistent with the previous observation in K${_3}$picene,\cite{picene2} which provides a good probe for determining the amount of doping electrons. The downshifts of the Raman modes are not the same with nominal value $x$ because of the different donations of alkali metal. The experimental frequencies for the Raman peaks originally at 409 and 1350 cm$^{-1}$ are plotted as a function of ${x}$ for K${_x}$phenanthrene in Fig. 3. These two Raman peaks roughly locate at three discrete values in nominal value $x$ from 0 to 6.0, $i.e.$ 393, 395, 399 cm$^{-1}$, and 1322, 1325, 1328 cm$^{-1}$. In the previous studies on K${_x}$picene\cite{picene2} and A${_x}$C$_{60}$, \cite{c1,c2} Raman scattering can be applied to determine the ${x}$ value of these compounds. Based on these analyses, there are two phases in K${_x}$phenanthrene which are K$_{3.0}$phenanthrene (1328 cm$^{-1}$) and K$_{4.0}$phenanthrene (1322 cm$^{-1}$). Besides, we found the other frequency (1325 cm$^{-1}$) intermediate between 1328 and 1322 cm$^{-1}$, which suggests the existence of K$_{3.5}$phenanthrene phase. Similar with the case of K${_x}$picene, only three discrete phases can be produced in K${_x}$phenanthrene which is independent of the nominal value ${x}$, and other phases with different composition of potassium can not be fabricated.

In these three possible phases, only the phase of K$_{3.0}$phenanthrene exhibits the superconducting transition based on the magnetization measurement. This obversion is consistent with the previous discovery of superconductivity\cite{XFWang} in phenanthrene that only the sample with nominal composition of K$_{3.0}$phenanthrene shows superconductivity in a series of  K${_x}$phenanthrene samples with different potassium contents, and all other samples with ${x}$ deviation from 3 do not show any feature of superconductivity. However, our results suggest that the superconductivity of K${_x}$phenanthrene is not based on the nominal value but closely correlates with actual amount of electrons transferred to the molecule. In the hydrocarbon based and carbon based superconductor, the donation of three electrons to a molecule plays an important role on the superconductivity, which is independent to the nominal donation of alkali metal.

\begin{figure}[tbp]
\includegraphics[width=\columnwidth]{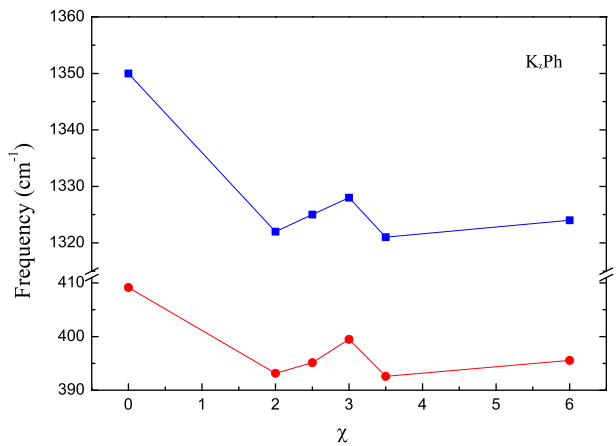}
\caption{(color online) Dependence of frequency of experimental Raman peaks (originally at 409 and 1350 cm$^{-1}$) on the nominal value ${x}$ in K${_x}$phenanthrene (0 $\leq$ ${x}$ $\leq$ 6).}
\end{figure}

Based on the analysis of the positions of Raman modes, we observe three phases in K${_x}$phenanthrene. In all samples, we only observe superconductivity in the phase (${x}$ = 3.0). The sample of other phases do not show any features of superconductivity. Additionally, the phases of K$_{3.5}$phenanthrene and K$_{4.0}$phenanthrene are much more stable than the superconducting phase (K$_{3.0}$phenanthrene). These two nonsuperconducting phases can be easily obtained in the sample of any nominal ${x}$ values. Thus, the superconductivity is pretty dependent on the amount of electrons transferred to the aromatic molecule, and is also very sensitive to the external conditions especially the annealing condition.

In order to investigate the equilibrium established in our superconducting K$_{3.0}$phenanthrene sample, we examined the the Raman-active modes of  K$_{3.0}$phenanthrene after further annealing process (shown in the Fig. 4) in comparison with the pristine phenanthrene. The blue line is the Raman spectra annealing at 473 K for 40 hours showing apparent superconducting transition at 5 K. Furthermore, we performed another annealing process at 453 K for 80 hours after the initial preparation. Based on the magnetization measurement, superconductivity of K$_{3.0}$phenanthrene disappears after the second annealing process. In the first annealing process, the most Raman modes shift to lower frequencies due to the intercalation of potassium atoms. After further annealing procedure, however, some Raman modes are different to those in the initial K$_{3.0}$phenanthrene. The original vibration modes at 409, 710, 1036, 1350  cm$^{-1}$ still remain but become weaker, and some new peaks occur in Raman spectra. We can see that the Raman spectra for K$_{3.0}$phenanthrene prepared by shorter annealing time, where the peak shift to lower frequency at 1328 cm$^{-1}$ from 1350 cm$^{-1}$ on pristine phenanthrene. On the other hand, the peak of K$_{3.0}$phenanthrene prepared by longer annealing time, shift to a frequency at 1357 cm$^{-1}$, showing that less electrons are transferred to the phenanthrene molecule from potassium atoms. Besides the Raman modes originating from phenanthrene molecule, some new peaks are observed at 360, 720 and 1270 cm$^{-1}$ in the sample prepared for longer annealing time, suggesting that the possible molecular arrangement or new material were produced in the further annealing process.

\begin{figure}[tbp]
\advance\leftskip-0.8cm
 \vspace{-12pt}
\includegraphics[width=1.15\columnwidth]{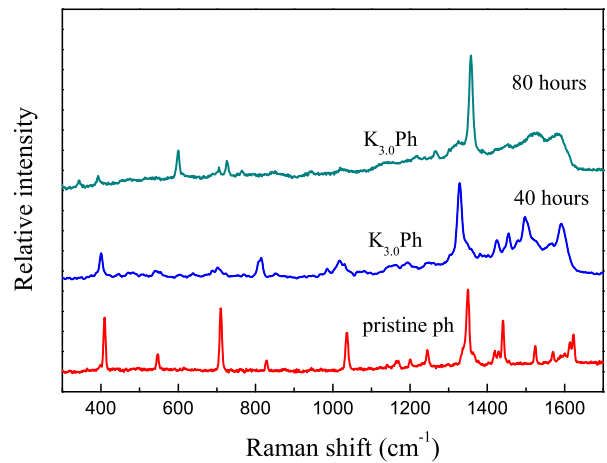}
\caption{(color online) Raman spectra of K$_{3.0}$phenanthrene respectively annealing at different time in comparison with pristine phenanthrene.}
\end{figure}

\begin{figure}[tbp]
\begin{center}
\advance\leftskip-2.5cm
 \vspace{-20pt}
\includegraphics[width=1.7\columnwidth]{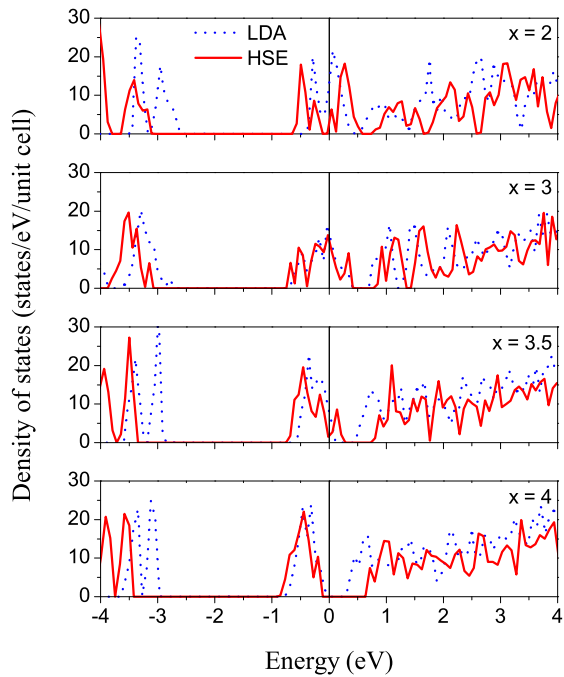}
\vspace{-20pt}
\caption{(color online) Calculated density of states of K${_x}$phenanthrene for $x=2, 3, 3.5,$ and 4. Dotted and solid lines are the LDA and HSE results, respectively.} 
\end{center}
\end{figure}

From the frequency range from 1500-1700 cm$^{-1}$, these vibration modes represent the C-C stretching modes. In the Raman spectra of K${_x}$phenanthrene (nominal ${x}$=2.0 2.5 6.0), most Raman modes representing the C-C stretching modes disappear or become featureless. The featureless in this region is possibly related to the molecular distortion of benzene rings and slightly molecular arrangement due to the doping of potassium metal.
This indicates that the rigid-band approximation is not valid in the calculations for hydrocarbons. By doping potassium metal and employing the density functional calculations, we have theoretically investigated the crystal and electronic structures of solid K${_x}$phenanthrene. The calculations are performed using the Vienna \emph{ab} \emph{initio} simulation package (VASP)\cite{vasp} based on the local density approximation (LDA).\cite{lda} Considering the effects of electronic correlations in potassium-doped PAHs,\cite{Giovannetti, Ruff} the hybrid functionals method (HSE)\cite{hse1,hse2,hse3} is adopted to overcome the limitations of LDA. Figure 5 shows the calculated density of states for four selected doping concentrations. From the results based on LDA (dotted lines), the solid K${_x}$phenanthrene exhibits the metallic feature at $x=2, 3$, and 3.5 while becomes to insulator at $x=4$. Noticeably, the density of states at Fermi level for the dopings of $x=2$ and $x=3.5$ are visibly less than that of $x=3$, which implies a decrease of superconducting electronic states when the $x$ deviates from 3. In fact, under the effect of electronic correlations, the Fermi level is located in the pseudo-gap of the electronic states of K${_2}$phenanthrene and K$_{3.5}$phenanthrene systems, as shown HSE results (solid lines) in Fig. 5. Consequentially, the metallic feature is absent in the systems where deviating from $x=3$, which also explains the disappearance of superconductivity in these materials.

In addition, we also find that the arrangement of phenanthrene molecule and K atomic positions will result in the variations of density of states at Fermi level. This indirectly indicates that the superconductivity in not only depends on the doping content of alkali metal but also depends on the molecular arrangement and the atomic position of alkali metal. However, the information of location of potassium atom is still opened and calls for the future structural determination especially from neutron diffraction for further exploring mechanisms of superconductivity. However, the doping constrain on superconductivity discovered here is important for understanding the mechanism of superconductivity al least in phenanthrene-based superconductors.

\section{conclusions}

In summary, we have performed Raman scattering measurements on K$_{x}$phenanthrene in order to identify the doping-superconductivity phase relation. We found that there exist three phases (${x}$ = 3, 3.5, 4) in K-doped phenanthrene. Only the phase (${x}$ = 3) was found to exhibit the superconductivity at $T_{\rm c}$ of 5 K. Consequently, the donation of three electrons to a molecule is essential for superconductivity, which is independent to the nominal donation of alkali metal. This experimental finding of the cation constrain on superconductivity is helpful for designing or synthezing new hydrocarbon superconductors having higher transition temperatures.

\begin{acknowledgments}
We would like to express our appreciation to Prof. Xian-Hui Chen at University of Science and Technology of China for his hospitality for allowing us to use his laboratory in sample synthesis. This work was supported as part of EFree, an Energy Frontier Research Center funded by the U.S. Department of Energy (DOE), Office of Science under
DE-SC0001057. The work done in China was supported by the Cultivation Fund of the Key Scientific and Technical Innovation Project Ministry of Education of China (No.708070), the Shenzhen Basic Research Grant (No. JC201105190880A), and the National Natural Science Foundation of China (No. 11274335).
\end{acknowledgments}


\begin{references}

\bibitem{XFWang} X. F. Wang, R. H. Liu, Z. Gui, Y. L. Xie, Y. J. Yan, J. J. Ying, X. G. Luo, and X. H. Chen, Nat. Commun. \textbf{2}, 507 (2011).

\bibitem{picene} R. Mitsuhashi, Y. Suzuki, Y. Yamanari, H. Mitamura, T. Kambe, N. Ikeda, H. Okamoto, A. Fujiwara, M. Yamaji, N. Kawasaki, Y. Maniwa, Y. Kubozono, Nature \textbf{464}, 76 (2010)

\bibitem{Kubozono} Y. Kubozono, H. Mitamura, X. Lee, X. He, Y. Yamanari, Y. Takahashi, Y. Suzuki, Y. Kaji, R. Eguchi, K. Akaike, T. Kambe, H. Okamoto, A. Fujiwara, T. Kato, T. Kosugi, and H. Aoki, Phys. Chem. Chem. Phys. \textbf{13}, 16476 (2011).

\bibitem{GFChen} M. Xue, T. Cao, D. Wang, Y. Wu, H. Yang, X. Dong, J. He, F. Li, and G. F. Chen, Scientific Reports \textbf{2}, 389 (2012).

\bibitem{TTF} A. M. Kini, U. Geiser, H. H. Wang,K. D. Carlson,J. M. Williams, W. K. Kwok, K. G. Vandervoort, J. E. Thompson, and D. L. Stupka, Inorg. Chem. \textbf{29}, 2555 (1990).

\bibitem{BEDTTF} H. Taniguchi, M. Miyashita, K. Uchiyama, K. Satoh, N. Mori, H. Okamoto, K. Miyagawa, K. Kanoda, M. Hedo and Y. Uwatoko, J. Phys. Soc. Jpn. \textbf{72}, 468 (2003)

\bibitem{TMTSF} D. Jerome, A. Mazaud, M. Hirano, and H. Hosono, J. Am. Chem. Soc. \textbf{130}, 3296 (2008).

\bibitem{C60} A. F. Hebard, M. J. Rosseinsky, R. C. Haddon, D. W. Murphy, S. H. Glarum, T. T. M. Palstra, A. P. Ramirez and A. R. Kortan, Nature \textbf{350}, 600, (1991).

\bibitem{Kato} T. Kato, K. Yoshizawa, and K. Hirao, J. Chem. Phys. \textbf{116}, 3420 (2002).

\bibitem{Kato2} T. Kato, T. Kambe, and Y. Kubozono, Phys. Rev. Lett. \textbf{107}, 077001 (2011).

\bibitem{Subedi} A. Subedi and L. Boeri, Phys. Rev. B \textbf{84}, 020508(R) (2011).

\bibitem{Casula} M. Casula, M. Calandra, G. Profeta, and F. Mauri Phys. Rev. Lett. \textbf{107}, 137006 (2011).

\bibitem{Giovannetti} G. Giovannetti and M. Capone, Phys. Rev. B \textbf{83}, 134508 (2011).

\bibitem{guo} G. H. Zhong, C. Zhang, G. F. Wu, Z. B. Huang, X. J. Chen, and H. Q. Lin, J. Appl. Phys. {\bf 113}, 17E131 (2013).

\bibitem{T1} T. Kosugi, T. Miyake, S. Ishibashi, R. Arita, and H. Aoki, J. Phys. Soc. Jpn. \textbf{78}, 113704 (2009).

\bibitem{T2} P. L. de Andres, A. Guijarro, and J. A. Verges, Phys. Rev. B \textbf{83}, 245113 (2011).

\bibitem{T3} T. Kosugi, T. Miyake, S. Ishibashi, R. Arita, and H. Aoki, Phys. Rev. B \textbf{84}, 214506 (2011).

\bibitem{PS} H. Okazaki, T. Wakita, T. Muro, Y. Kaji, X. Lee, H. Mitamura, N. Kawasaki, Y. Kubozono, Y. Yamanari, T. Kambe, T. Kato, M. Hirai, Y. Muraoka, and T. Yokoya, Phys. Rev. B \textbf{82}, 195114 (2010).

\bibitem{huang} Q. W. Huang, J. Zhang, A. Berlie, Z. X. Qin, X. M. Zhao, J. B. Zhang, L. Y. Tang, J. Liu, C. Zhang, G. H. Zhong, H. Q. Lin, and X. J. Chen, J. Chem. Phys. {\bf 139}, 104302 (2013).

\bibitem{zhao} X. M. Zhao, J. Zhang, A. Berlie, Z. X. Qin, Q. W. Huang, J. Shan, J. B. Zhang, L. Y. Tang, J. Liu,  C. Zhang, G. H. Zhong, H. Q. Lin, and X. J. Chen, J. Chem. Phys. {\bf
139}, 144308 (2013).

\bibitem{fane} S. Fanetti, M. Citroni, L. Malavasi, G. A. Artioli, P. Postorino, and R. Bini, J. Phys. Chem. C {\bf 117}, 5343 (2013).

\bibitem{capi} F. Capitani, M. H\"{o}ppner, B. Joseph, L. Malavasi, G. A. Artioli, L. Baldassarre, A. Perucchi, M. Piccinini, S. Lupi, P. Dore, L. Boeri, and P. Postorino, Phys. Rev. B {\bf 88}, 144303 (2013).

\bibitem{c1} T. Pichler, M. Matus, J. Kurti, and H. Kuzmany, Phys. Rev. B \textbf{45}, 13841 (1992).

\bibitem{c2} S. Fujiki, Y. Kubozono, S. Emura, Y. Takabayashi, S. Kashino, A. Fujiwara, K. Ishii, H. Suematsu, Y. Murakami, Y. Iwasa, T. Mitani, and H. Ogata, Phys. Rev. B \textbf{62}, 5366 (2000).

\bibitem{picene2} T. Kambe, X. He, Y. Takahashi, Y. Yamanari, K. Teranishi, H. Mitamura, S. Shibasaki, K. Tomita, R. Eguchi, H. Goto, Y. Takabayashi, T. Kato, A. Fujiwara, T. Kariyado, H. Aoki, and Y. Kubozono, Phys. Rev. B \textbf{86}, 214507 (2012).

\bibitem{vasp}
G. Kresse and J. Furthmuller, Comput. Mater. Sci. \textbf{6}, 15 (1996).

\bibitem{lda}
J. P. Perdew and Y. Wang, Phys. Rev. B \textbf{45}, 13244 (1992).

\bibitem{Ruff}
A. Ruff, M. Sing, R. Claessen, H. Lee, M. Tomi\'{c}, H. O. Jeschke, and R. Valent\'{i}, Phys. Rev. Lett. \textbf{110}, 216403 (2013).

\bibitem{hse1}
J. Heyd, G. E. Scuseria, and M. Ernzerhof, J. Chem. Phys. \textbf{118}, 8207 (2003).

\bibitem{hse2}
J. Heyd and G. E. Scuseria, J. Chem. Phys. \textbf{121}, 1187 (2004).

\bibitem{hse3}
J. Heyd, G. E. Scuseria, and M. Ernzerhof, J. Chem. Phys. \textbf{124}, 219906 (2006).


\end{references}
\end{document}